\documentclass[11pt,letter]{article}
\usepackage{hyperref}
\pdfoutput=1

\begin{document}

\title{Chimney Formation in Mushy Layers -- Experiment and Simulation}
\author{
Anthony M.~Anderson$^1$, Richard F.~Katz$^2$, and Grae Worster$^1$ \\
\\\vspace{6pt} $^1$Department of Applied Math and Theoretical Physics, \\
University of Cambridge, Cambridge, UK \\
\\\vspace{6pt} $^2$Department of Earth Sciences, \\
University of Oxford, Oxford, UK
}
\date{20--22 November, 2011}

\maketitle

\begin{abstract}
In this fluid dyanmics video, we show experimental images and simulations 
of chimney formation in mushy layers. A directional solidification apparatus
was used to freeze 25 wt\% aqueous ammonium chloride solutions at controlled
rates in a narrow Hele-Shaw cell (1mm gap). The convective motion is imaged
with schlieren. We demonstrate the ability to numerically simulate mushy
layer growth for direct comparison with experiments.
\end{abstract}

Mushy layers come into being during the rapid solidification of binary liquid 
melts. They are composed of aligned dendritic crystals that are bathed in residual 
liquid. This residual interstitial liquid is typically of a different composition 
than the bulk liquid, which can trigger convection within the mushy layer. When 
convection is present, dissolution often leads to the formation of chimneys -- drainage 
channels devoid of solid. Chimneys are the source of defects known as freckles 
in industrial casting processes. Chimneys are also conduits for brine drainage 
from young sea ice. A major goal in the study of mushy layers is to predict the 
formation of chimneys and the amount of solute transported through them.

We have used a directional solidification apparatus to conduct experiments on 
aqueous ammonium chloride (NH$_{4}$ Cl) at fixed freezing rates in a Hele-Shaw cell 
with a 1 mm gap thickness. Since the solution concentrations used were larger than 
the eutectic concentration and freezing was from below, the interstitial fluid 
in the mushy layer was lighter and fresher than the surrounding fluid, producing 
double-diffusive convection. Using schlieren, we imaged the density field to reveal 
the convective fluid motion and chimney formation for a range of freezing rates 
and starting concentrations. An experimental video is shown for a 25 wt\% solution
frozen at 1 $\mu$m/s. 

A mushy layer can be modeled as a reactive porous matrix whose porosity changes 
in order to ensure local thermodynamic equilibrium. The use of a Hele-Shaw cell 
with a narrow gap in the above experiments greatly simplifies the modelling of 
mushy layers since Darcy's Law can be used throughout the entire domain. Katz and 
Worster (J. Comp. Phys, 2008) developed a 2D numerical model based on this 
simplification and the Enthalpy method. With this approach, the prescription of 
internal boundary conditions at the solid, mush, and liquid interfaces could be 
avoided. With these simulations, we have the opportunity to make direct comparisons
with experiments. This capability is demonstrated in a simulation video using 
the same conditions as in the experimental video. 

\end{document}